\documentclass[usenatbib]{mn2e}
\usepackage{graphics}
\usepackage{epsfig}
\usepackage{natbib}
\usepackage{color}

\begin{document}

\title[Clustering of MgII absorption line systems]
{Clustering of MgII absorption line systems around massive
galaxies: an important  constraint on feedback processes 
in galaxy formation}

\author [G.Kauffmann et al.] {Guinevere Kauffmann$^1$\thanks{E-mail: gamk@mpa-garching.mpg.de},
Dylan Nelson$^1$, Brice M\'enard$^2$, Guangtun Zhu$^2$\\
$^1$Max-Planck Institut f\"{u}r Astrophysik, 85741 Garching, Germany\\
$^2$Department of Physics \& Astronomy, Johns Hopkins University,
Baltimore, MD, 21218, USA}

\maketitle

%===================================
\begin{abstract} 
We use the latest version of the metal line absorption catalogue of Zhu \&
M\'enard (2013) to study the clustering of MgII absorbers around massive
galaxies ($\sim 10^{11.5} M_{\odot}$), quasars and radio-loud AGN with
redshifts between 0.4 and 0.75.  Clustering is evaluated in two dimensions,
by binning absorbers both in projected radius and in velocity separation.
Excess MgII is detected around massive galaxies out to $R_p=20$ Mpc. At
projected radii less than 800 kpc, the excess extends out to velocity
separations of 10,000 km s$^{-1}$.  The extent of the high velocity tail
within this radius is independent of the mean stellar age of the galaxy and
whether or not it harbours an active galactic nucleus.  We interpret our
results using the publicly available Illustris and Millennium simulations.
Models where the MgII absorbers trace the dark matter particle or subhalo
distributions do not  fit the data. They overpredict the clustering on
small scales and do not reproduce the excess high velocity separation MgII
absorbers seen within the virial radius of the halo.  The Illustris
simulations which include thermal, but not mechanical feedback from AGN,
also do not provide an adequate fit to the properties of the cool halo gas
within the virial radius. We propose that the large velocity separation
MgII absorbers trace gas that has been pushed out of the dark matter halos,
possibly by multiple episodes of AGN-driven mechanical feedback acting over
long timescales.
\end{abstract}


\section{Introduction}
An improved understanding of the distribution of gas in dark matter halos
is key to solving the so-called ``overcooling'' problem in galaxy
formation, which is based on the deduction that if gas simply cools over a
Hubble time in dark matter halos in a universe dominated by cold dark
matter, too many very high mass galaxies will be produced (White \& Frenk
1991; Kauffmann, White \& Guiderdoni 1993).

An important diagnostic of the ability of gas to cool comes from X-ray
observations of groups and clusters. The deepest observations with the
Chandra satellite now allow measurements of the gas mass fractions of
groups as far out as $r_{500}$ and it is found that the lowest mass systems
have the lowest hot gas fractions (e.g. Sun et al 2009). One question has
been to what extent these results are affected by the fact that the groups
and clusters with deep X-ray observations are almost always X-ray selected.
In recent work, Anderson et al (2015) stacked a sample of 250,000 locally
brightest galaxies from the Sloan Digital Sky Survey using data from the
ROSAT All-Sky Survey, which was used to derive a relation between galaxy
mass and average X-ray luminosity. Wang et al (2016) then used weak
gravitational lensing to measure the total mass profiles around the same
sample of stacked galaxies, thus allowing a derivation of the relation
between X-ray luminosity and galaxy halo mass.  The results yield a scaling
between $L_x$ and $M_{500}$ that has a similar slope compared to previous
work, but with a normalization that is a factor of 2 below the relations derived
from X-ray selected samples.

These results suggest that there is considerable variation in the baryon
content of halos at fixed galaxy mass.  They also suggest that gas heating
mechanisms play an important role in determining the {\em global}
thermodynamic state of the gas in lower mass halos.  McCarthy et al (2010)
analyzed the predicted X-ray properties of X-ray groups in cosmological
hydrodynamical simulations from the OverWhelmingly Large Simulations (OWLS)
project and showed that the models with AGN feedback produced X-ray scaling
relations  in much better agreement with observations than the models where
the only gas heating mechanism is from supernovae. So far, however, X-ray
observations are unable to constrain the physical nature of the heating
process.  Choi et al (2014,2015) compare two different AGN feedback models:
one that includes only thermal heating, and another that includes
mechanical feedback. The mechanical model was motivated by observations of
winds in broad absorption line quasars (BAL QSOs), which convey energy,
mass and momentum into the surrounding gas with velocity $\sim 10,000$ km s$^{-1}$
(see Crenshaw, Kraemer \& George 2003 for a review). Choi et al (2015) show that for a fixed
amount of energy released during a given black hole accretion event, the
mechanical model is much more efficient than the thermal model at pushing
gas out of the central region of the halo and reducing the predicted X-ray
luminosities. In the simulation with mechanical feedback , gas particles
are given fixed kick velocities of 10,000 km s$^{-1}$, suggesting that the kinematics
of the gas in groups and clusters might also be an important test of
models.

The Mg II $\lambda \lambda$2796,2803 absorption doublet observed in quasar
spectra traces low-ionization gas over a broad range in HI column density
($10^{16.5} < N_{HI} < 10^{21.5}$ cm$^{-2}$). Since the early exploratory
work in the late 1980's (Bergeron 1986; Sargent, Steidel  \&
Boksenberg 1988), Mg II has evolved into a useful probe of the gaseous halos
around galaxies thanks to very large samples of MgII systems that have been
extracted from quasar spectra observed as part of the Sloan Digital Sky
Survey (Lundgren et al 2009; Quider et al 2011; Zhu \& Menard 2013;
P{\'e}rez-R{\`a}fols et al 2015).  There have been many studies of the spatial
clustering of Mg II systems around massive galaxies observed as part of the
Baryon Oscillation Spectroscopic Survey (BOSS; Dawson et al. 2013), which
have median stellar masses $\sim 10^{11.5} M_{\odot}$. Of particular
note is the study of Zhu et al (2014), where the average cool gas surface
density profile traced by MgII systems around massive galaxies was measured
out to projected radii of 10 Mpc.  A change of slope is observed on scales
of 1 Mpc, consistent with the expected transition from the regime where the
gas is typically within the same dark matter halo as the galaxy, to the
regime where the gas is outside the parent halo.  Zhu et al introduce a
model where the gas distribution is assumed to trace the dark matter
distribution exactly, i.e. the gas density profile in a halo of mass M has
the same NFW shape as the dark matter up to a normalization factor
$f_{gas}(M_{halo})$.  This model is shown to provide an adequate fit to
the data if the average host halo mass of the galaxies is $10^{13.5}
M_{\odot}$.

In principle, more stringent constraints on the distribution of gas within
dark matter halos can be obtained if the clustering amplitude of MgII
systems around galaxies is studied both as a function of projected radius
$R_p$ and as a function of velocity separation $\Delta V$. Wild et al
(2008) were the first to study both the transverse and the line-of-sight clustering of  a large
sample of many thousands of CIV and MgII 
absorption line systems around quasars using data from the SDSS data
release 3. These authors found  a non-zero correlation between CIV systems
and quasars in the redshift interval $0.4<z<2.2$, extending out
to line-of-sight velocity separations of 12,000 km s$^{-1}$, which could
not be explained by assuming that these systems were associated with the
foreground galaxy population. They proposed that the high velocity CIV
systems instead represented an {\em outflowing gas component} that was
directly associated with the quasar itself. There was also evidence that
high velocity CIV absorbers were more numerous around radio-loud quasars
compared to radio-quiet quasars.  The sample of MgII absorbers was much
smaller than the sample of CIV absorbers, so although there was also a hint
of an excess high velocity component of gas traced by MgII, it could not be
as accurately quantified.

In this paper, we make use of the galaxy, quasar and quasar absorption line
catalogues from the SDSS DR12 (the final data release of the BOSS survey)
to compare the line-of-sight clustering of MgII absorbers around massive
galaxies, quasars and radio-loud AGN in bins of projected radius from
$R_p=100$ kpc to $R_p=15$ Mpc over the redshift interval 0.4 to 0.75.  We
interpret our results using 
full cosmological N-body simulations of dark matter plus gas from the
publicly available Illustris simulation (Nelson et al 2015). The N-body+gas
simulations incorporate recipes for how AGN heat the gas in and around
galaxies. Our paper is structured as follows. In Section 2, we decribe the
data sets, the construction of our catalogues, and the methodology used for
computing the clustering signal. Our empirical results are presented in
Section 3 and Section 4 attempts to further interpret these results using 
simulations.

\section{Samples and Methods}

\subsection {The galaxy sample} The sample of massive galaxies used in this
work originates from the twelfth data release (DR12) of the SDSS. Spectra
of about 1.5 million galaxies are available from the Baryonic Oscillation
Spectroscopic Survey (BOSS; Dawson et al 2013).  We select a subsample of
galaxies that meet the CMASS target selection criterion, which was designed
to select an approximately stellar mass limited sample at $z>0.45$. The
sample is defined using the following cuts: 
\begin {eqnarray}
17.5 < i < 19.9 & d_{\bot}> 0.55 &    \\ 
i < 19.86+1.6(d_{\bot}-0.8) &  r-i < 2 &  i_{fibre} < 21.5
\end{eqnarray}
where $d_{\bot}$ is a `rotated' combination of colours defined as
$d_{\bot}= (r-i)-(g-r)/8$. This constraint identifies galaxies that lie at
high enough redshift so that the 4000 \AA\ break has shifted beyond the
observer frame $r$-band. $i_{fibre}$ is the amount of light that enters the
fibre.  We note that all colour cuts are defined using ``model'' magnitudes,
whereas magnitude limits are given in terms of ``cmodel'' magnitudes.

We extract a sample of 937,079 galaxies that satisfy the CMASS sample
selection cuts and that have redshifts in the range $0.4<z<0.75$ (the lower 
limit is fixed by the detectability of  MgII systems in SDSS
quasar spectra and the upper limit by the magnitude limit of 
the CMASS galaxies (Zhu \& M\'enard 2013)). In this paper we work with stellar
masses and estimates of the 4000 \AA\ break index derived using methods
based on principal component analyses (see Chen et al 2012 for more
details).  These methods are designed to maximize the information content
in low $S/N$ spectra and they have been shown to reproduce stellar mass
estimates based on broad-band photometry and direct measurements of the
4000 \AA\ break in high $S/N$ spectra of SDSS galaxies at low redshifts.  
The distribution of stellar masses and 4000 \AA\ break strengths of the
galaxies in our sample is shown in the top two panels of Figure 1. The
galaxies in our sample lie in the stellar 
mass range $10^{11}< M_*<10^{12} M_{\odot}$ and
the median mass is around $3 \times 10^{11} M_{\odot}$.  The distribution
of 4000 \AA\ break strengths indicates that the galaxies are predominantly
composed of old stars --  a single burst of star formation at very high
redshift followed by passive evolution yields a 4000 \AA\ break strength of
1.7 at this redshift (assuming solar metallicity), 
while constant star formation over a Hubble time
yields a 4000 \AA\ break strength of 1.4. Significant
tails of galaxies with young stars are, however, seen for galaxies with stellar
masses closer to $10^{11} M_{\odot}$ (lower left panel). As discussed in
Chen et al (2012), the fraction of massive galaxies 
($\log M_* > 11.2$) with active star formation evolves strongly with redshift,
decreasing  by a
factor of $\sim 5$ from  $z=0.6$ to $z=0.1$.

\begin{figure}
\includegraphics[width=91mm]{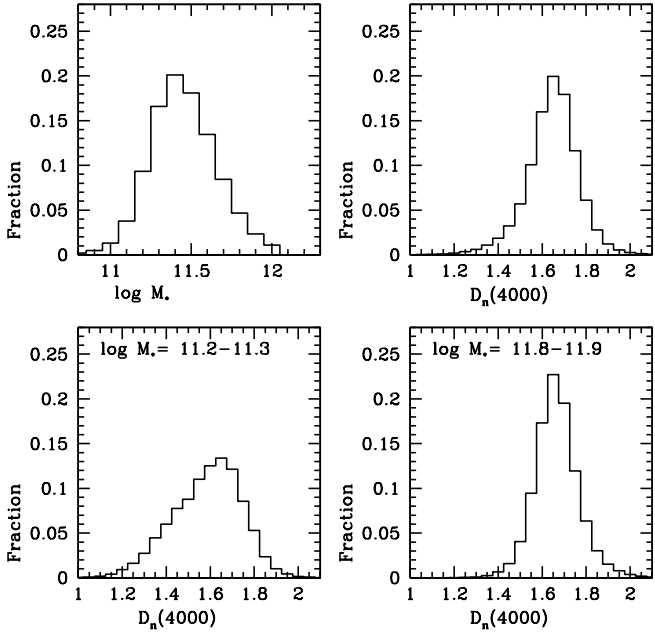}
\caption{ {\em Top left:} Stellar mass distribution of the galaxies in our sample. 
{\em Top right:} Estimated 4000 \AA\ break strength distribution of the galaxies.
{\em Bottom panels:} Estimated 4000 \AA\ break strength distributions in two
stellar mass ranges. 
\label{models}}
\end{figure}

\subsection {The MgII absorber sample} We begin with a catalogue of 52,243
MgII absorbers extracted from 142,012 quasar spectra from the DR12. Of these, 
11035 MgII absorbers have redshifts that overlap the CMASS galaxy sample. The
details of the absorption-line detection algorithm are given in Zhu \&
Menard (2013). The algorithm consists of the following steps:
\begin {enumerate}
\item  continuum
estimation using a basis set of eigenspectra \item  filtering out fluctuations
on intermediate scales \item a search using a multi-line model that includes
MgII $\lambda \lambda$ 2796, 2803 and 4 strong FeII lines: $\lambda$ 2344,
$\lambda$ 2382, $\lambda$ 2586 and $\lambda$ 2600.  The inclusion of the Fe II
lines eliminates the fraction of false positives by reducing confusion in
the case of quasars with multiple absorption-line systems. \item  
double-Gaussian profile fitting of the MgII $\lambda\lambda$ 2796, 2803 doublet.
The dispersions of the two Gaussians are assumed to be identical, but their
centers and amplitudes are free parameters in the fitting algorithm.
Candidates are rejected if the separation beween the Gaussians differs from
the fiducial value by more than 1 \AA\ \item  Estimation of the absorber
redshift and the rest-frame equivalent widths of the lines.
\end {enumerate}

Zhu \& Menard (2013) carry out tests of the completeness of their detection
algorithm by adding simulated absorbers drawn from a distribution of rest
equivalent widths and doublet ratios. The completeness is higher for
stronger absorbers and at redshifts for which the noise level of the flux
residuals is lower.  Low completeness spikes are found in the region of
prominent sky lines and in the wavelength region where data from the red
and blue arms of the spectrograph are joined. These completeness estimates
are used to correct the observed numbers of absorbers and to derive the
differential incidence rate $dN/dz$ as a function of rest equivalent width.
The incidence rate of weaker absorbers with $0.6 < W_0^{\lambda 2796} <
1.0$ \AA\ is consistent with constant co-moving density. The incidence of
stronger absorbers increases by a factor $\sim 3$ out to redshifts of 2,
before dropping.

\subsection {Random galaxy catalogues} The goal of the analysis is to
quantify the clustering amplitude and spatial extent of the MgII absorbers
associated with massive galaxies as a function of projected radius
$R_p$ and velocity separation $\Delta V$. In similar fashion to the
analysis in Wild et al (2008), we implement a method for computing the
excess number of galaxy-absorber pairs relative to an unclustered sample of
galaxies.

This is  done by computing the number of absorbers expected at random along
the actual sightlines used to find the observed pairs. To do this, we
create random galaxy samples by randomizing the positions of the galaxies
on the sky, but also making sure that the galaxies lie within the
boundaries of the CMASS galaxy sample. The redshift of each object 
is drawn at random from the redshift distribution of the
CMASS sample, i.e. the galaxies in the random catalogue and the CMASS
sample have exactly the same redshift distributions. This procedure
accounts for the evolving number density of MgII absorbers with redshift and
the fact that the detection completeness varies as a function of
wavelength.  The procedure is also straightforward to implement in N-body
simulations, as we discuss in Section 4. We have not implemented detailed
survey masks, i.e.  we have not accounted for the fact that there may be
areas within the boundaries of the survey where galaxies are not targeted
due to the presence of bright stars, asteroid trails and other imaging
imperfections. As we will show, this may introduce a spurious excess
clustering signal of $\sim 10\%$ on the largest scales, but does not affect
our primary conclusions.

\section {Observational Results}

We first carry out a check to test the robustness of our procedure for
estimating the number of randomly associated absorbers along a given
sightline, by plotting the  clustering signal separately for absorbers with
positive and negative velocity separations with respect to their parent
galaxies. Results are shown in Figure 2. We plot the logarithm of the
number of galaxy-absorber pairs in the CMASS sample divided by the average
number of galaxy-absorber pairs in the random catalogues as a function of
velocity separation. Results are shown in 6 different bins in projected
radius $R_p$, ranging from 160 kpc to 15 Mpc.

\begin{figure}
\includegraphics[width=91mm]{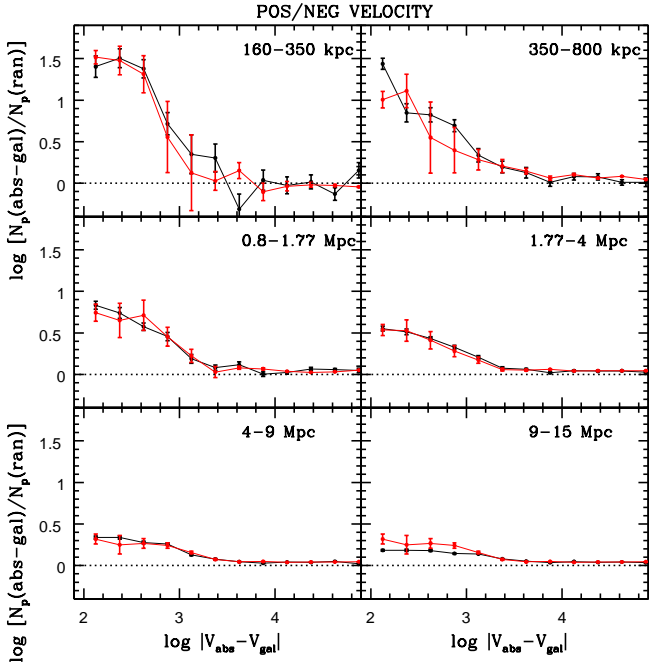}
\caption{ The logarithm of the
number of galaxy-absorber pairs in the CMASS sample divided by the average
number of galaxy-absorber pairs in the random catalogues is plotted as a function of
velocity separation. Results are shown in 6 different bins in projected
radius $R_p$. Results for absorbers with negative velocity deparations are plotted
in black, and for positive velocity separations in red.  
\label{models}}
\end{figure}

As can be seen, a positive clustering signal is detected out to 15 Mpc, and
out to velocity separations well beyond 1000 km s$^{-1}$. Error bars have been
estimated via  bootstrap resampling.  We generate 30 different random CMASS
catalogues, so the error budget is  dominated by the small number of close
separation quasar-galaxy pairs in the real CMASS catalogue. We find that
the clustering signals in the positive and negative velocity directions are
consistent within the errors. This gives us confidence that our procedure
for normalizing out the ``background'' absorbers is giving robust results.
There is a small (0.05 dex) residual positive clustering signal at large
velocity separations that represents the overall systematic error in the
construction of our random catalogues.

We then combine the clustering estimates for positive and negative velocity
separations to produce the result shown in Figure 3 (provided in tabular
form in Table 1). Excess clustering is
clearly detected out to velocity separations of $\sim$ 10,000 km s$^{-1}$ in the
bins with projected radii less than 800 kpc. At large projected radius,
excess absorbers are seen out to velocity separations of 6000 km s$^{-1}$. The
clustering amplitude at velocity separations less than 1000 km s$^{-1}$ is still
nearly a factor of two in excess of the background in the $R_p=9-15$ Mpc
bin.

\begin{figure}
\includegraphics[width=91mm]{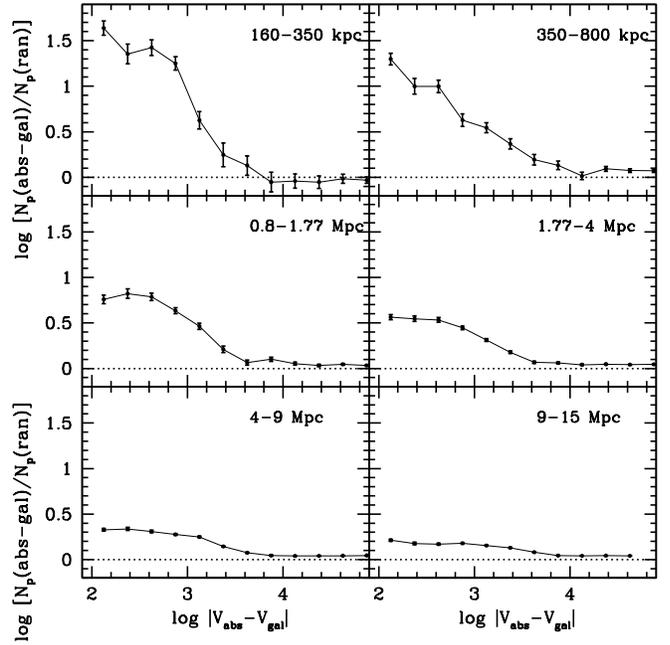}
\caption{ The logarithm of the
number of galaxy-absorber pairs in the full CMASS sample divided by the average
number of galaxy-absorber pairs in the random catalogues is plotted as a function of
the absolute value of the velocity separation. Results are shown in 6 different bins in projected
radius $R_p$. 
\label{models}}
\end{figure}

\begin{table*}
\centering
\caption{The logarithm of the
number of galaxy-absorber pairs in the full CMASS sample divided by the average
number of galaxy-absorber pairs in 30 random samples is tabulated  as a function of
velocity separation for the six bins in projected radius in Figure 3}
\begin{tabular}{lcccccc}
log $\Delta$V &  0.16-0.35 Mpc & 0.35-0.8 Mpc & 0.8-1.77 Mpc & 1.77-4 Mpc & 4-9 Mpc  & 9-15 Mpc \\
2.125 &  1.638$\pm0.078$ & 1.299$\pm0.064$ & 0.760$\pm0.047$ & 0.563$\pm0.029$ & 0.328$\pm0.015$ & 0.213$\pm0.011$ \\  
2.375 &  1.354$\pm0.108$ & 0.999$\pm0.087$ & 0.821$\pm0.052$ & 0.546$\pm0.030$ & 0.337$\pm0.017$ & 0.177$\pm0.014$ \\
2.625 &  1.424$\pm0.087$ & 0.999$\pm0.069$ & 0.786$\pm0.041$ & 0.533$\pm0.027$ & 0.309$\pm0.015$ & 0.170$\pm0.010$ \\
2.875 &  1.250$\pm0.074$ & 0.628$\pm0.068$ & 0.634$\pm0.033$ & 0.446$\pm0.022$ & 0.276$\pm0.010$ & 0.179$\pm0.008$ \\ 
3.125 &  0.626$\pm0.096$ & 0.544$\pm0.054$ & 0.464$\pm0.034$ & 0.312$\pm0.016$ & 0.249$\pm0.009$ & 0.155$\pm0.005$ \\
3.375 &  0.247$\pm0.130$ & 0.367$\pm0.056$ & 0.210$\pm0.035$ & 0.178$\pm0.014$ & 0.144$\pm0.006$ & 0.130$\pm0.005$ \\
3.625 &  0.130$\pm0.107$ & 0.193$\pm0.058$ & 0.063$\pm0.028$ & 0.068$\pm0.012$ & 0.076$\pm0.006$ & 0.081$\pm0.003$ \\
3.875 & -0.053$\pm0.108$ & 0.132$\pm0.046$ & 0.101$\pm0.023$ & 0.060$\pm0.009$ & 0.045$\pm0.004$ & 0.045$\pm0.003$ \\
4.125 & -0.042$\pm0.077$ & 0.015$\pm0.040$ & 0.053$\pm0.018$ & 0.040$\pm0.007$ & 0.040$\pm0.004$ & 0.042$\pm0.002$ \\  
4.375 & -0.052$\pm0.069$ & 0.092$\pm0.027$ & 0.033$\pm0.013$ & 0.047$\pm0.006$ & 0.041$\pm0.003$ & 0.044$\pm0.002$ \\ 
4.625 & -0.016$\pm0.049$ & 0.074$\pm0.021$ & 0.046$\pm0.010$ & 0.042$\pm0.005$ & 0.042$\pm0.002$ & 0.042$\pm0.001$ \\
4.875 & -0.032$\pm0.035$ & 0.072$\pm0.017$ & 0.034$\pm0.009$ & 0.045$\pm0.003$ & 0.046$\pm0.002$ & 0.042$\pm0.001$ \\
\end{tabular}
\end{table*}

In Figures 4 and 5, we examine whether the clustering signal shows
dependence on galaxy mass or on the age of the stellar population as
measured by the 4000 \AA\ break strength. The CMASS sample is divided into
two equal high/low mass subsamples at log $M_* \sim 11.5$ and  into
red/blue subsamples at a  4000 \AA\ break strength of 1.7. As can be seen a
dependence on galaxy properties is largely absent. MgII absorvers are
slightly more numerous around blue galaxies compared to red galaxies in the
smallest bin in projected radius (160-350 kpc).  Lan, Menard \& Zhu (2014)
have used photomeric catalogues from SDSS and the Galaxy Evolution Explorer
(GALEX) satellite to extract a much larger sample of galaxy-absorber close pairs
spanning a wide range in galaxy colour. They also find that differences in MgII
covering fractions only become significant at small projected radii and
most of their analysis is restricted to galaxy-absorber pairs with
separations less than 50 kpc. There are only a handful of galaxy-absorber pairs
in our sample with such small separations. Note also that the galaxy
sample spans a very limited range in stellar mass ($10^{11}-10^{12} M_{\odot}$)
where the relation between galaxy mass and halo mass is very shallow, so it
is perhaps not surprising that there is little dependence
of the clustering signal on stellar mass.

\begin{figure}
\includegraphics[width=91mm]{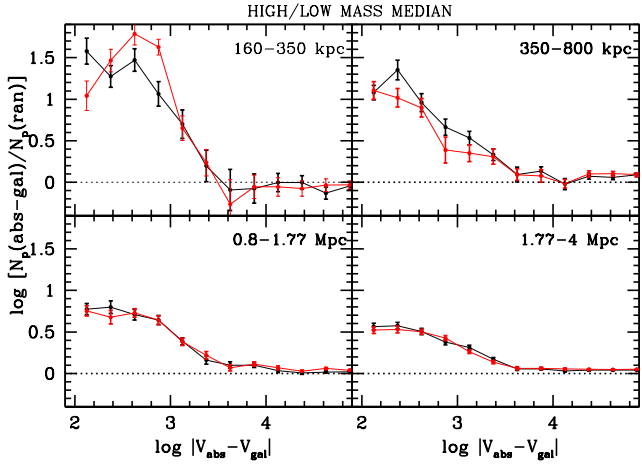}
\caption{ The logarithm of the
number of galaxy-absorber pairs in the full CMASS sample divided by the average
number of galaxy-absorber pairs  in the random catalogues is plotted as a function of
velocity separation. Results are shown for galaxies with $\log M_* < 11.5$ in black
and for $\log M_* > 11.5$ in red.  
\label{models}}
\end{figure}

\begin{figure}
\includegraphics[width=91mm]{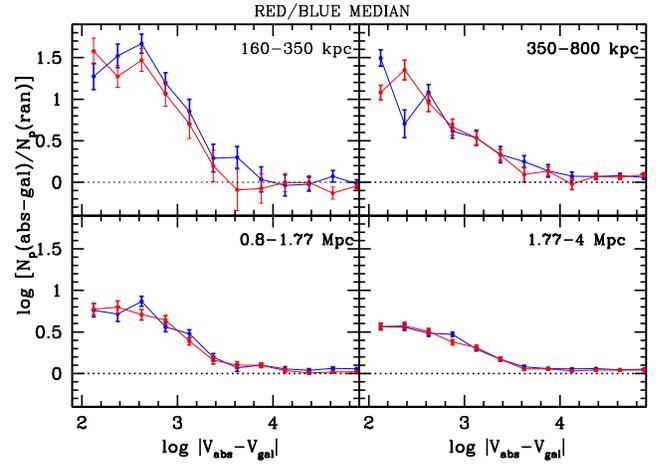}
\caption{ The logarithm of the
number of galaxy-absorber pairs in the full CMASS sample divided by the average
number of galaxy-absorber pairs in the random catalogues is plotted as a function of
velocity separation. Results are shown for galaxies with 4000 \AA\ break
index strength less than 1.7 in blue and greater than 1.7 in red.  
\label{models}}
\end{figure}

Finally, in order to make contact with the analysis of Wild et al (2008),
we have carried out similar clustering analysis of MgII absorbers around
SDSS quasars. We have also examined clustering of Mg II 
around galaxies with radio-loud  nuclei  in the
same redshift range ($0.4<z<0.75$) as the CMASS sample. \footnote {Note
that the Wild et al analysis is for radio-loud quasars rather than galaxies}. To find radio-loud
galaxies, we cross-match the CMASS sample with the source catalog from the Faint
Images of the Radio Sky at Twenty-Centimeters  (FIRST) survey carried out
at the VLA (Condon et al. 1998). The SDSS and FIRST positions are required to be within 3
arcseconds of each other. Because radio loud nuclei are known to be biased
towards the very most massive galaxies (Best et al 2005), we also extract a
`control' sample of galaxies with the same stellar mass distribution that
are selected without regard to their radio properties.  We note that we do
not impose a flux limit on our radio loud galaxy sample as we wish to maximize
the number of sources included in the clustering analysis. Inclusion of
false detections will act to dilute clustering differences compared to the
control sample.  There are 43,812 galaxies in the radio loud galaxy  sample.
We also select a sample of 18,338 quasars in the same redshift interval
from the SDSS DR12 quasar catalogue (P\^aris et al. 2016).  
Because the light from the central nucleus outshines the underlying host galaxy
in quasars, we are unable for construct a sample of normal galaxies matched
in stellar mass for this sample.

In the left-hand panels in Figure 6, we compare the clustering of MgII
absorbers around radio-loud galaxies  with the control sample of galaxies matched
in stellar mass. As can be seen, the distribution of MgII systems as a
function of velocity separation is the same around radio-loud galaxies  as it is
around normal galaxies.  Both samples exhibit  clear excess of MgII
absorbers out to velocity separations of $\sim 10,000$ km s$^{-1}$.  It is the
{\em amplitude} of the clustering signal of MgII absorbers that is higher
around radio-loud galaxies, indicating that there is more cool gas around these
systems. In the right hand panel, we compare the clustering of MgII
absorbers around radio-loud galaxies and quasars.  The results for the quasar
sample are quite a lot noisier because of the smaller sample size. We find
no clear differences in the distribution of MgII gas around radio-loud
galaxies and quasars.

\begin{figure}
\includegraphics[width=91mm]{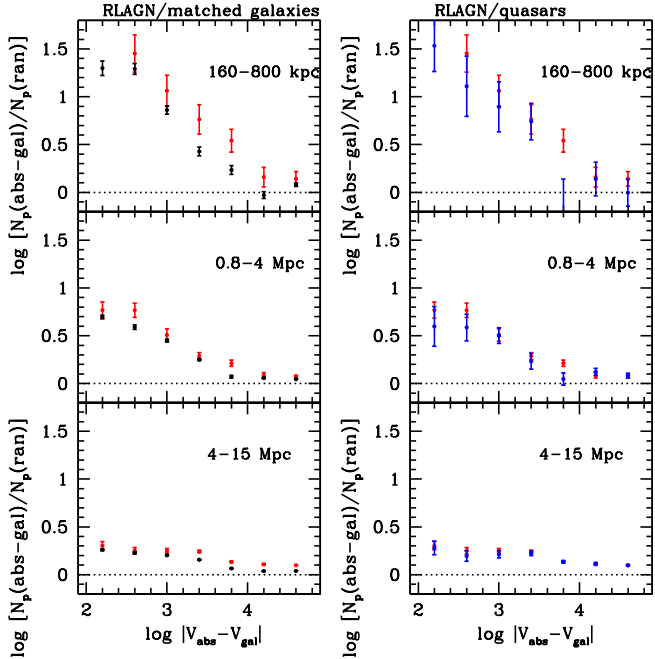}
\caption{ {\em Left panels:} the line-of-sight clustering of MgII
absorbers around radio-loud galaxies (red) is compared  with the control sample of galaxies matched
in stellar mass (black). Results are shown for 3 bins
in projected separation. {\em Right panels:} The clustering of MgII
absorbers around radio-loud galaxies  (red) and quasars (blue). 
\label{models}}
\end{figure}

Our examination of differences in the clustering of MgII systems around
galaxies with different 4000 \AA\ break strengths and around quasars and
radio-loud galaxies  do not support a scenario in which 
the  high velocity gas components are intrinsic to an active galactic
nucleus that is is instantaneously ejecting gas at very high velocity. If so,
one might expect that no high velocity gas components  be found around
passive galaxies with no AGN. As we will discuss in the next section, our results 
suggest that the high-velocity MgII absorbers trace gas that has been {\em
pushed out} of galactic halos at some time in the past. This hypothesis is
also supported by the fact that the 
high velocity separation absorbers  are not confined to the
bins with small projected separation from the parent galaxy, but are seen
out to projected radii of tens of Mpc. We come back to this point in the
final section.

\section {Models}

In this section, we attempt some simple interpretation of our results with
the help of cosmological simulations 
that also incorporate gas dynamics. A full accounting of
observational selection effects is beyond the scope of this paper. To do
this, we would need to generate mock quasar absorption line spectra from
the simulations and take into account the observed MgII detection thresholds in
the data (see Hernquist et al 1996 for discussion of how to generate mock
quasar spectra and Ford et al 2016 for discussion of how 
mock quasar spectra from simulations can be used in 
the interpretation of  metal absorption
line data from the COS-HALOS survey).

The approach we take is similar to that of Zhu et al (2014), except we use
simulations rather than analytic models to predict the clustering signal,
given a set of possible assumptions for how the MgII gas might trace the
underlying large scale structure dominated by cold dark matter. The main
advantage of simulations over analytic models is that we can account for
large-scale structure present {\em outside} the virial radius of the dark
matter halo. This is necessary in order to address the question of whether
high velocity tails of MgII absorbers can be explained in a simple way, or
whether we need to invoke models in which AGN feedback significantly alters
the structure and kinematics of the gas.

We make use of simulation data released by the Illustris and Millennium cosmological
simulation projects. The reader is referred to
http://www.illustris-project.org/ and Nelson et al (2015) for detailed
information about the Illustris project and to http://wwwmpa.mpa-garching.mpg.de/millennium/
and Lemson et al. (2006) for information about the Millennium simulation.  

We first show results using  dark matter particle and
gas data from the Illustris-3 simulation and dark matter subhalo data from
the Illustris-1 simulation at an output time that corresponds to the median
redshift of the CMASS sample ($z=0.55$, snapshot 101).   All the simulations are within a  periodic
box of length 106.5 Mpc. The dark matter particle mass in the Illustris-3
simulation is $\sim 4 \times 10^8 M_{\odot}$.  The average gas cell mass is
$8.1 \times 10^7 M_{\odot}$. The Illustris-1 simulation has the same box
size, but is run with a dark matter particle mass of $7.5 \times 10^6
M_{\odot}$. The Illustris-3 simulation is used to investigate models where
the MgII absorber population is assumed to trace the dark matter or gas
distributions. The Illustris-1 simulation is used to investigate models
where the MgII absorber population traces dark matter subhalos. The higher
resolution is needed so that we can track subhalos down to low masses
within groups and clusters.  In low resolution simulations, low mass
subhalos are tidally disrupted more easily.

We investigate the following simple tracer models
and show comparisons with the observational data in Figure 7: \begin{enumerate} \item
absorbers trace the dark matter particle distribution (red curves in Figure
7) \item absorbers trace dark matter subhalos down to some limiting mass
(dotted  magenta and red curves in Figure 7 show results for limiting
masses of $10^{7.5}$ and $10^{8.5} M_{\odot}$, respectively) \item
absorbers trace the total gas distribution (blue curves in Figure 7)
\item absorbers trace the neutral gas distribution (cyan curves in
Figure 7). Neutral gas fractions are read directly from the Illustris
simulation. \item absorbers  trace gas with $T < 10^5$ K (green curves in
Figure 7), for which the MgII mass fraction is high (see below).  \end {enumerate}

In order to mimic the observational set-up, we select galaxies from the
Illustris-1 and Illustris-3 simulations with stellar masses in the range
$10^{11}-10^{12} M_{\odot}$. There are 150 such galaxies in the
Illustris-3 simulation and 369 in Illustris-1. The larger number of massive
galaxies in Illustris-1 is the result of enhanced cooling of gas in high
density regions that are more highly resolved. As discussed in Vogelsberger
et al (2014), there is an excess of massive galaxies compared to
observations in Illustris-1 of about a factor of two at stellar masses of
$\sim 10^{11.5} M_{\odot}$.

Results for all the Illustris tracer models are shown in Figure 7. The models where
absorbers trace the dark matter, the dark matter subhalo population or the
total gas distribution all lie above the observations at small velocity
separations ($\Delta V < 500$ km/s), and then fall below the data at
$\Delta V > 1000$ km/s. The model where absorbers trace the neutral gas
distribution shows no enhancement of neutral gas around massive galaxies
compared to random sightlines. This is because the galaxies in our sample
typically reside in dark matter halos of masses $\sim 10^{13} M_{\odot}$ or
greater, where most of the gas in the halo has been shock-heated to high
enough temperatures to ionize almost all the hydrogen. In addition,
``radio-mode" AGN feedback following Sijacki et al (2007) where bubbles of
hot gas with radius 50 kpc, total energy $10^{50}$ erg are placed within
the halo, also acts to stop gas from cooling in halos of this mass.

\begin{figure}
\includegraphics[width=91mm]{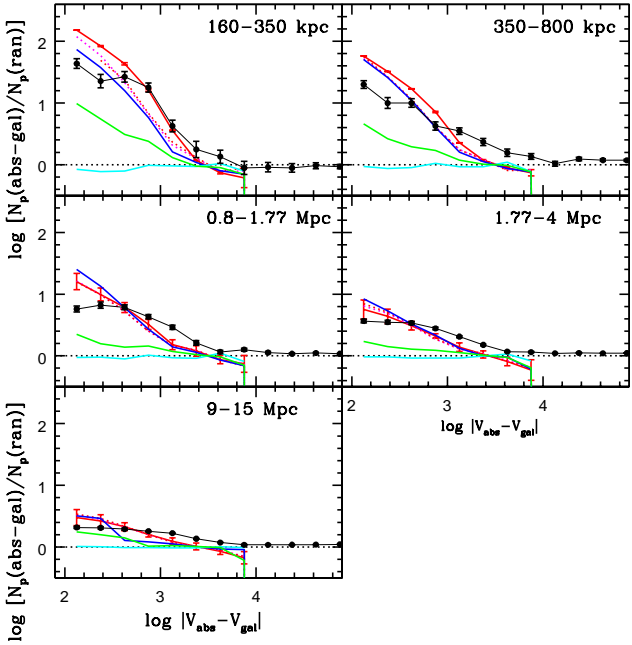}
\caption{ Line-of-sight clustering predictions for different tracer models are compared to
the observational data. Please see the text for details of the different models.
Black curves show the same data as in Figure 3. Solid red curves are for a model
where the absorbers trace the dark matter particle distribution. Dotted red and magenta
curves are models where the absorbers trace the dark matter subhalo population.
Cyan curves are for a model where the absorbers trace the neutral gas component.
Green curves are for a model where the absorbers trace gas with temperature $T < 10^5$ K.   
\label{models}}
\end{figure}

The tracer model that comes closest to mimicking how MgII absorbers might be
distributed in the Illustris simulation is the model  where absorbers trace
gas with $T < 10^5$ K.  This is illustrated in Figure 8, which shows, as a
function of density and temperature, where the dominant fractions of
Magnesium are in the form of Mg II. To compute these ionic abundances we
use CLOUDY (v13.03, Ferland et al. 2013) including both collisional and
photo-ionization in the presence of a UV background (Ferland 2009). We follow
Bird et al. (2015) and use CLOUDY in single-zone mode, accounting for a
frequency dependent shielding from the background radiation field at high
densities, using the fitting function of Rahmati et al. (2013). Under these
assumptions MgII is found predominantly in gas with densities above
$10^{-4}$ cm$^{-3}$ and temperatures below $10^5$ K. The MgII fractions are
also roughly constant in this regime. As can be seen from Figure 7 (green
curves in the plot), in this model, MgII absorbers are significantly more
clustered around massive galaxies than in random sightlines, but the
amplitude is lower than in the observations. In addition, there is no tail
to large velocity separations as in the data.

In order to evaluate the clustering out to
velocity separations of 10,000 km/s in the Illustris simulation, we
have replicated the simulation box
periodically in the x,y and z directions.  We caution that missing large
scale power will result in an under-prediction of the clustering amplitude
at large velocity separations comparable to the box size. In addition, the integral
constraint on the correlation function also implies that clustering at large
velocity separations will be under-estimated. In order to investigate
this, we compare results for the Illustris tracer model (i) with results obtained
for the Millennium simulation, which has a box size  500$h^{-1}$ Mpc, i.e. 6.7  times larger
than the Illustris box. Once again, we select galaxies with stellar masses
in the range $10^{11}-10^{12} M_{\odot}$ from the $z=0.5$ snapshot to represent the
galaxy population. We note that unlike Illustris, the parameters of the galaxy
formation models in the Millennium simulations are tuned to provide
an accurate fit to the galaxy luminosity function. In particuar, the
volume density of massive galaxies predicted by the Millennium simulation
is in much better agreement with observational constraints than in Illustris (Croton et al 2006).  

Figure 9 shows that clustering amplitude predicted by the Millennium dark matter
model is significantly higher than that predicted by Illustris. As we have
discussed, the Illustris simulation overproduces massive galaxies and they
thus will trace dark matter halos with lower masses, leading to a lower
clustering amplitude prediction. The Millennium dark matter
model also converges cleanly to zero overdensity at large velocity separations,
indicating that box size is now sufficiently large. 
The model provides a better match to the clustering amplitude
at large $\Delta$V in the bins with large projected radius.
We see that there is still 
a significant tail of high velocity MgII absorbers in the 350-800 kpc bin.  
The interpretation of the size of the excess is compromised by the
fact that the Millennium model overshoots the clustering amplitude at small $\Delta$V by
such a large factor. 
  
In summary, none of the models explored in this section provide an adequate fit
to the observational data. Models in which the MgII absorbers trace the dark matter
are too highly clustered at small velocity separations. This is in agreement with
the conclusions of Wild et al (2008). There appears to be a significant high velocity
tail of absorbers that is especially apparent at projected radii between 300 -800 kpc,
which cannot be reproduced by any of the models explored in this paper.
It will be interesting to see whether the large  $\Delta$V problem will  be resolved in  
models with black hole driven thermal and kinetic feedback 
(e.g. Weinberger et al 2016). It also remains to be seen whether the MgII absorber population
predicted by applying CLOUDY photo-ionization models to simulations that better resolve
the small scale structure of the gas, can provide a better fit at small $\Delta$V. 

\begin{figure}
\includegraphics[width=94mm]{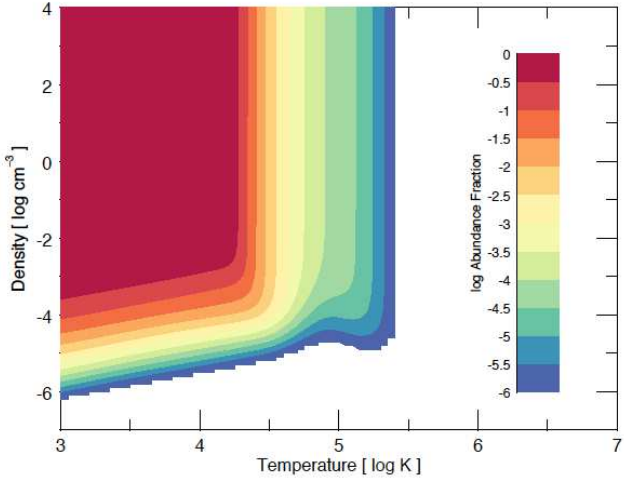}
\caption{ The fraction of
Magnesium in the form of Mg II as a function of density and temperature as 
predicted by  CLOUDY in single-zone mode, including both collisional and
photo-ionization in the presence of a UV background (see text). 
\label{models}}
\end{figure}

\begin{figure}
\includegraphics[width=91mm]{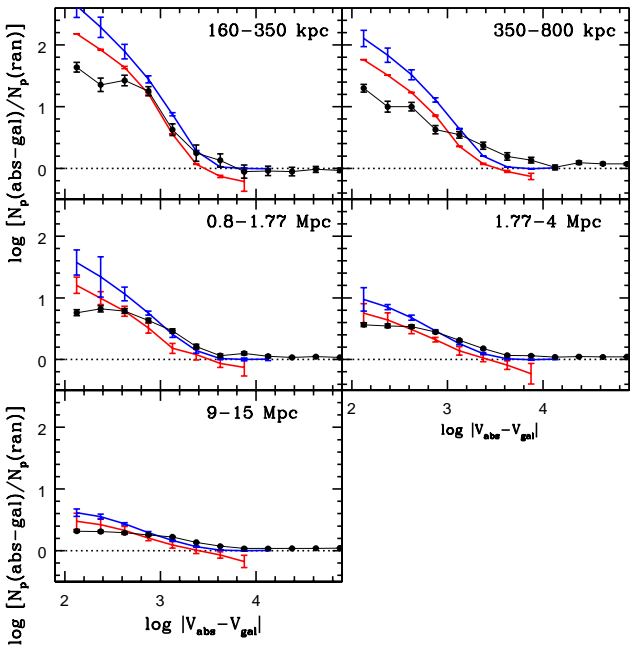}
\caption{ Line-of-sight clustering predictions for the dark matter tracer model
in the Illustris simulation (red) is compared to that for the Millennium simulation (blue).
The observational data are shown as black points. 
\label{models}}
\end{figure}

\section {Discussion}

In section 4, we showed that none of our tracer models were able to account
for the clustering signal at large velocity separations from the  parent
galaxy. In section 3, we proposed that the large $\Delta V$ absorber
population was unlikely to be associated with outflowing material from an
active galactic nucleus, because it is detected out to large (at least 1 Mpc) projected
radius $R_p$. We note that Wild et al. (2008) and all other previous work on
associated absorbers only investigated line-of-sight clustering of CIV and MgII
absorbers to the host QSO, and only probed the high velocity gas in front
of the QSO. Wild et al. additionally measured the transverse clustering of the
absorbers around the QSOs, but did not investigate the velocity distribution of this
gas. 

It is also interesting to investigate  how far out in projected radius we
are able to detect a clustering signal.
In Figure 10, we  show 
the number of absorber-CMASS galaxy pairs divided by the number of
absorber-random galaxy pairs is plotted in 3 bins of projected radius: 4-9
Mpc, 9-20 Mpc, and 20-45 Mpc. Results are shown for 4 different bins in
stellar mass $M_*$. Note that the y-axis in this plot is in linear rather
than logarithmic units, in order to  see the extent of the low amplitude
clustering signal more clearly. A significant excess of MgII absorbers is
detected out to velocities well beyond 3000 km/s even in the 9-20 Mpc bin.
The excess clustering has largely disappeared at all velocity separations
in the 20-45 Mpc bin.

\begin{figure*}
\includegraphics[width=130mm]{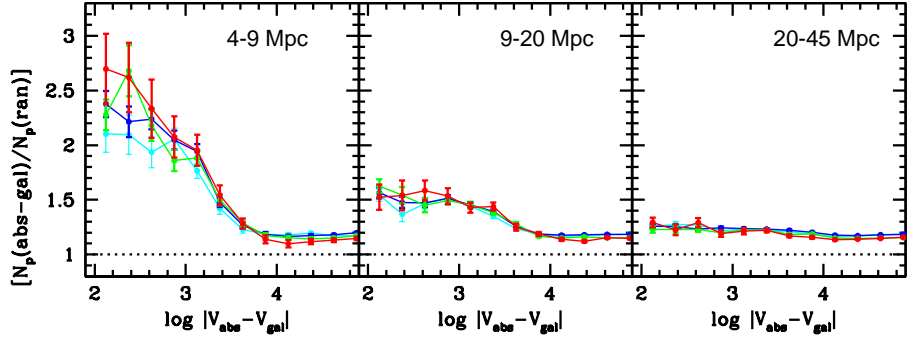}
\caption{ The 
number of galaxy-absorber pairs divided by the average
number of galaxy-absorber pairs in the random catalogues is plotted as a function of
velocity separation. Results are shown in 3 different bins in projected
radius $R_p$ from 4 to 45 Mpc. Different colour curves show results for different
stellar mass intervals: $10^{11}-10^{11.2} M_{\odot}$ (cyan), 
$10^{11.2}-10^{11.4} M_{\odot}$ (blue), $10^{11.4}-10^{11.6} M_{\odot}$ (green),
$10^{11.6}-10^{11.8} M_{\odot}$ (red). 
\label{models}}
\end{figure*}

One possible  interpretation of our results is that the
MgII absorbers with large $\Delta V$ trace gas that has been pushed out of
dark matter halos.  A series of papers (Kauffmann et al 2013; Kauffmann
2015) have investigated an interesting large-scale clustering phenomenon
that we now propose may be closely
related. In brief, it has been found  that the correlation between the
colours and specific star formation rates of neighbouring galaxies  first
noticed by Weinmann et al (2006) and dubbed ``galactic conformity'', extends
far beyond the scale of the virial radii of their dark matter halos. In
particular, Kauffmann (2015) showed that low mass galaxies with low
specific star formation rates are surrounded by neighbouring galaxies with
lower than average specific star formation rates out to projected radii of
$\sim 20$ Mpc, which is very similar to the scales over which we find the
excess MgII absorber population.  The interstellar medium of low mass
galaxies travelling through large-scale reservoir of gas with temperatures
of $10^5-10^6$ K may be stripped by ram-pressure forces, leading to a
shut-down in star formation that is correlated over scales of many Mpc, as
seen in the data.

Kauffmann (2015) also find a significant excess of very high mass ($\log
M_* > 11.3$) galaxies around low $SFR/M_*$ central galaxies and an even
larger excess of high mass galaxies hosting radio-loud AGN around these
systems. This again suggests a link between the large-scale MgII absorber
excess and the galactic conformity phenomenon. As shown in Figure 6, the
number of MgII systems around radio-loud AGN is significantly higher than in
control samples of the same stellar mass, which probably implies that the
total gas density is higher in the vicinity of radio-loud AGN and that
ram-pressure stripping effects on low mass galaxies will be more pronounced.

The final question we address is what physical process is responsible for
pushing gas out of galactic halos. We do not find any correlation between
the presence of a population of large velocity separation absorbers and a
radio-loud active galactic nucleus. This does not rule out the possibility
that  excess gas on large scales may have been pushed out of the halo by
{\em multiple episodes} of AGN feedback over timescales of many Gyr.
Tremonti et al (2007) obtained  spectra of 14 massive galaxies with  $M_*
\sim 10^{11} M_{\odot}$ and redshifts similar to the CMASS galaxies in our
sample. In 10/14 cases, the MgII $\lambda 2796, 2803$ absorption lines were
blue-shifted by 500-2000 km/s with respect to the stars. Chen et al (2013)
have carried out an analysis of the star formation histories of CMASS
galaxies and find that the spectra of  galaxies with ongoing star formation
are best fit by models that include recent bursts. This motivates us to search for a 
a sub-population of MgII systems at small impact
parameters that have been ejected more recently. 

Two properties of the MgII absorber
systems are available from the catalogue of Zhu \& M\'enard (2013): the absorption
line equivalent width and the doublet ratio $W_0^{2796}/W_0^{2803}$.  We
extract all MgII absorbers within a projected radius of 500 kpc from a
CMASS galaxy, and we plot the doublet ratio and the equivalent width of
the $\lambda 2796$ MgII line as a function of velocity separation in the
left-hand panels of Figure 10.  In the right-hand panels, we plot doublet
ratio and equivalent width as a function of $R_p$ for all absorbers within
$\Delta V < 10,000$ km/s. The red error bars indicate the running median of
the distribution, while the green error bars indicate the upper 75th
percentile.  There is no trend in either equivalent width or doublet ratio
as a function of $R_p$ for the population of absorbers as a whole. We note,
however, that we cannot extend our analysis to $R_p$ smaller than 100 kpc,
because of poor sample statistics.
The top left panel of Figure 11  provides an interesting hint that the
population of high velocity MgII absorbers located at small impact
parameters may be systematically different from the absorbers associated
within the galaxy's halo.  There is a decrease in doublet ratio as a
function of $\Delta V$ from  $\Delta V = 50$ km/s out to $\Delta V= 700$
km/s, followed by a flat relation out to very large velocity separations.
Note that  $\Delta V= 700$ km/s corresponds reasonably well to the boundary
expected for gas clouds that are in virial equilibrium within  dark
matter halos in the mass range $10^{13}-10^{14} M_{\odot}$. 
The $W_0^{2796}/W_0^{2803}$ ratio is expected to be bounded
between 2 (the optically thin regime) and 1 (saturated). If the low $\Delta
V$ MgII absorbers correspond to gas clouds that are in virial
equilibrium within the inner regions of the halo, they may be more
optically thin because ram-pressure and tidal forces, as well as
collisional heating processes, have acted to reduce their column densities.

\begin{figure}
\includegraphics[width=94mm]{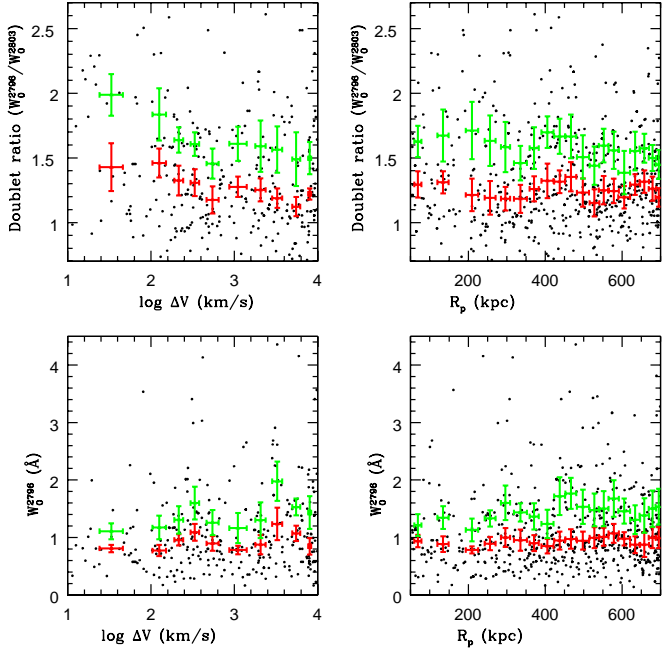}
\caption{ {\em Left panels:} The doublet ratio $W_0^{2796}/W_0^{2803}$ and the equivalent width
$W_0^{2796}$ is plotted as a function of velocity separation for absorbers with
$R_p < 500$ kpc. Red points with error bars show estimates of the running median.
Green points with error bars show the upper 75th percentile points. 
Error bars have been computed using a standard bootstrap resampling method.
{\em Right panels:} The same two quantities are plotted as a function of
projected radius $R_p$ for absorbers with $\Delta V < 10,000$ km s$^{-1}$.
\label{models}}
\end{figure}

In future work, we plan to analyze MgII $\lambda 2796, 2803$ absorption in
the actual spectra of CMASS galaxies and see whether we can find more
robust evidence for 
outflowing systems.
The number of MgII systems observed at small impact parameters is extremely
limited in this analysis because both the quasar and the CMASS galaxy
samples have low densities on the sky. In future, it will be valuable to
increase the coverage in the vicinity of high mass galaxies by targeting
background quasars down to fainter limits.  Spectroscopic surveys of
galaxies in the same redshift range to fainter limiting magnitudes would
allow us to study the kinematics of the gaseous halos of galaxies spanning
a wide range in stellar mass.  Higher resolution, higher
signal-to-noise spectra would be valuable in order to understand the
physical conditions in the gas traced by the MgII systems in more detail.
Finally, more detailed modelling of the MgII
absorber population is clearly needed to understand the implications of
the observational results in greater depth.

%===================================
\section*{Acknowledgments}
We thank Simon White and Philipp Busch for helpful input to the analysis.

%===================================

%===================================

\end{document}